\documentclass[iop,apj]{emulateapj}
\usepackage{apjfonts}
\usepackage{hyperref}

\journalinfo{The Astrophysical Journal}
\slugcomment{Received 2015 November 30; accepted 2016 February 3}

\shortauthors{CAMILO ET AL.}
\shorttitle{MILLISECOND PULSAR IN 3FGL~J1417.5$-$4402}

\begin{document}

\def\fermi{{\em Fermi}}
\def\swift{{\em Swift}}
\def\chandra{{\em Chandra}}
\def\msun{$M_{\sun}$}
\def\rsun{$R_{\sun}$}

\def\psr{PSR~J1417$-$4402}
\def\fgl{3FGL~J1417.5$-$4402}

\title{Discovery of a millisecond pulsar in the 5.4~day binary
3FGL~J1417.5$-$4402: observing the late phase of pulsar recycling
}

\author{F.~Camilo\altaffilmark{1,2},
  J.~E.~Reynolds\altaffilmark{3},
  S.~M.~Ransom\altaffilmark{4},
  J.~P.~Halpern\altaffilmark{1},
  S.~Bogdanov\altaffilmark{1},
  M.~Kerr\altaffilmark{3},
  P.~S.~Ray\altaffilmark{5},
  J.~M.~Cordes\altaffilmark{6},
  J.~Sarkissian\altaffilmark{7},
  E.~D.~Barr\altaffilmark{8},
  and E.~C.~Ferrara\altaffilmark{9}
}

\altaffiltext{1}{Columbia Astrophysics Laboratory, Columbia University,
  New York, NY~10027, USA}
\altaffiltext{2}{SKA South Africa, Pinelands, 7405, South Africa}
\altaffiltext{3}{CSIRO Astronomy and Space Science, Australia
  Telescope National Facility, Epping, NSW~1710, Australia}
\altaffiltext{4}{National Radio Astronomy Observatory, Charlottesville,
  VA~22903, USA}
\altaffiltext{5}{Space Science Division, Naval Research Laboratory,
  Washington, DC~20375-5352, USA}
\altaffiltext{6}{Department of Astronomy and Center for Radiophysics
  and Space Research, Cornell University, Ithaca, NY~14853, USA}
\altaffiltext{7}{CSIRO Parkes Observatory, Parkes, NSW~2870, Australia}
\altaffiltext{8}{Centre for Astrophysics and Supercomputing, Swinburne
  University of Technology, Hawthorn, VIC~3122, Australia}
\altaffiltext{9}{NASA Goddard Space Flight Center, Greenbelt,
  MD~20771, USA}

\begin{abstract}
In a search of the unidentified \fermi\ gamma-ray source \fgl\ with
the Parkes radio telescope, we discovered \psr, a 2.66\,ms pulsar
having the same 5.4 day orbital period as the optical and X-ray
binary identified by Strader et al. The existence of radio pulsations
implies that the neutron star is currently not accreting. Substantial
outflows from the companion render the radio pulsar undetectable
for more than half of the orbit, and may contribute to the observed
H$\alpha$ emission. Our initial pulsar observations, together with
the optically inferred orbit and inclination, imply a mass ratio
of $0.171\pm0.002$, a companion mass of $M_2=0.33\pm0.03$\,\msun,
and a neutron star mass in the range $1.77\leq M_1\leq2.13$\,\msun.
However, there remains a discrepancy between the distance of 4.4\,kpc
inferred from the optical properties of the companion and the smaller
radio dispersion measure distance of 1.6\,kpc. The smaller distance
would reduce the inferred Roche-lobe filling factor, increase the
inferred inclination angle, and decrease the masses. As a wide
binary, \psr\ differs from the radio-eclipsing black widow and
redback pulsars being discovered in large numbers by \fermi. It is
probably a system that began mass transfer onto the neutron star
after the companion star left the main sequence. The companion
should end its evolution as a He white dwarf in a 6--20 day orbit,
i.e., as a typical binary millisecond pulsar companion.

\end{abstract}


\section{Introduction} \label{sec:intro} 

The Large Area Telescope \citep[LAT;][]{aaa+09m} on the {\em Fermi
Gamma-ray Space Telescope\/} has been used to detect 200 rotation-powered
pulsars
\citep[e.g.,][]{aaa+13}\footnote{\url{https://confluence.slac.stanford.edu/display/GLAMCOG/Public+List+of+LAT-Detected+Gamma-Ray+Pulsars}}.
Nearly half of them are millisecond pulsars \cite[MSPs; e.g.,][]{aaa+09f}.
Half of the LAT-detected MSPs were discovered prior to \fermi\ in
``all-sky'' radio surveys. The other half were discovered in directed
radio searches of LAT unidentified sources \citep[e.g.,][]{rap+12}.

These two subpopulations differ in at least one important respect:
95\% of the MSPs discovered in all-sky surveys have white dwarf
companions or are isolated. By contrast, 40\% of those discovered
in LAT-guided searches are in compact ($<1$ day) interacting binaries
--- the so-called ``black widow'' or ``redback'' systems.  Black
widow pulsars have degenerate $\approx 0.03$\,\msun\ companions and
many are regularly eclipsed at radio wavelengths near pulsar superior
conjunction.  Redback MSPs have hot and bloated main sequence-like
$\ga 0.1$\,\msun\ companions and display irregular radio eclipses
that can persist for more than half of the orbit.

Black widows have long been considered to possibly represent the
final stage in the creation of isolated MSPs, but many uncertainties
remain \citep[e.g.,][]{prp02b}. How redbacks fit into the recycling
scenario of MSP formation, and how they may relate to black widows,
is even more uncertain \citep[e.g.,][]{ccth13,bdh14}.  No redbacks
were known in the Galactic disk prior to \fermi, and they continue
to present unexpected behavior: e.g., three redbacks have transitioned
rapidly between accretion-disk states with no radio pulsations and
rotation-powered states with radio pulsations \citep{asr+09,pdb+12,sah+14}.

In this context, \citet{scc+15} have discovered a peculiar system.
They report on multi-wavelength observations of the unidentified
source now known as \fgl, finding a variable X-ray object associated
with an optical counterpart that displays ellipsoidal variations
with a $P_b=5.4$ day binary period. They interpret the observation
of double-peaked H$\alpha$ emission as evidence for an accretion
disk in a low-mass X-ray binary, and conclude that the optical/IR
star is a $\approx 0.35$\,\msun\ giant at a distance of 4.4\,kpc
and that the neutron star is massive, $\approx 2.0$\,\msun.
\citet{scc+15} proposed that \fgl\ could be a transitioning MSP
that might eventually be detectable as a radio pulsar.  However,
the large orbit sets \psr\ apart from Galactic redbacks: all those
previously known have $P_b < 1$ day.

Here we report the discovery and initial study of a radio MSP that
is clearly the neutron star in the 5.4 day binary \fgl. Although
it shares some of the properties of redbacks, such as their companion
masses and extensive radio eclipses, \psr\ is on a standard
evolutionary track that will end as a typical binary MSP with a
low-mass white dwarf companion, as \citet{scc+15} deduced. It is
the first such system to be identified in the Galaxy.

\section{Observations} \label{sec:obs} 

\subsection{Radio Searches at Parkes} \label{sec:search} 

In \citet{ckr+15} we reported on a radio survey of 56 unidentified
LAT sources using the CSIRO Parkes telescope. We discovered 10 MSPs,
in some cases only after multiple observations of each source.
Multiple observations increase the chance that an otherwise detectable
pulsar will not be missed due to interstellar scintillation, large
acceleration at a particularly unfavorable binary phase, or eclipse.
We established that 23 of the remaining unidentified sources had
gamma-ray properties consistent with those of known pulsars, and
therefore were deserving of further radio searches.

During 2015 March--May we performed 28 search observations of 12
of the highest-ranked LAT targets remaining from our original source
list (Table~\ref{tab:search}). Because the analog filterbank used
in the original survey had in the meantime been decommissioned, we
used instead a digital filterbank (PDFB4, see \citealt{mhb+13}) to
record data from the central beam of the Parkes 20\,cm multibeam
receiver, at a center frequency of 1369\,MHz.  Each of 512 0.5\,MHz-wide
polarization-summed channels was sampled with PDFB4 every $80\,\mu$s
for approximately 1\,hr (see Table~\ref{tab:search}), and the
resulting power measures were written with 2-bit precision to disk
for off-line analysis.

\begin{deluxetable}{lcc}
\tablecaption{\label{tab:search} New Radio Searches of 3FGL Sources at Parkes }
\tablecolumns{3}
\tablehead{
\colhead{Source Name\tablenotemark{a}}   &
\colhead{Integration Time}               &
\colhead{$N_{\rm obs}$\tablenotemark{b}} \\
\colhead{}                               &
\colhead{(minutes)}                      &
\colhead{}
}
\startdata
 3FGL~J0940.6$-$7609  & 60, 60, 60, 41 & 5 \\
 3FGL~J1025.1$-$6507  & 60, 60, 60, 45 & 5 \\
 3FGL~J1231.6$-$5113  & 60, 80, 98, 52 & 5 \\
 3FGL~J1325.2$-$5411  & 60             & 2 \\
 3FGL~J1417.5$-$4402  & 60             & 2 \\
 3FGL~J1753.6$-$4447  & 60             & 2 \\
 3FGL~J1803.3$-$6706  & 55             & 2 \\
 3FGL~J1831.6$-$6503  & 60             & 2 \\
 3FGL~J2043.8$-$4801  & 45             & 3 \\
 3FGL~J2131.1$-$6625  & 60, 60, 60     & 5 \\
 3FGL~J2200.0$-$6930  & 55, 60, 60     & 4 \\
 3FGL~J2333.0$-$5525  & 60, 60, 60, 38 & 6
\enddata
\tablenotetext{a}{Source properties, including the positions that
we targeted in our searches, are given in \citet{aaa+15}.  See also
\citet{ckr+15}. }
\tablenotetext{b}{Number of radio searches done, including those
listed here and in \citet{ckr+15}. }
\end{deluxetable}

\begin{figure}
\begin{center}
\includegraphics[scale=0.72]{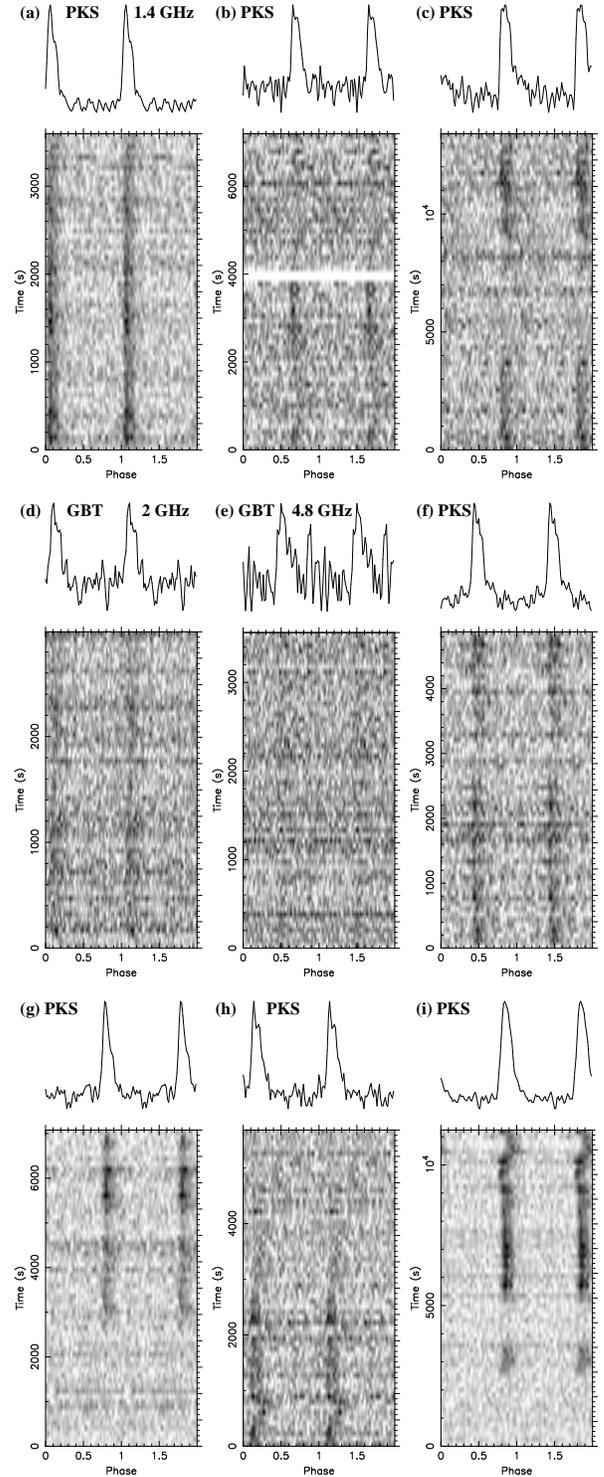}
\end{center}
\caption{\label{fig:profs}
All radio detections of \psr\ reported in this paper, in time order
(see also Figure~\ref{fig:dets}). All Parkes detections are at
1.4\,GHz. Individual integration times range between 1\,hr and
4\,hr. The folded pulse profiles are shown twice as a function of
time and summed at the top.
}
\end{figure}

We analyzed the data using PRESTO \citep{ran01}, exactly as for our
earlier searches described in \citet{ckr+15}. In an observation of
\fgl\ done on 2015 March 28, we discovered a binary pulsar with
period $P=2.66$\,ms and dispersion measure $\mbox{DM}=55$\,pc\,cm$^{-3}$
(Figure~\ref{fig:profs}a).  To our knowledge, this is the first
pulsar discovered using any of the PDFBs. We then realized that
\fgl\ was the gamma-ray source in which \citet{scc+15} had discovered
an unusual binary, and sought to determine whether the new pulsar
was related.

\subsection{Timing Observations of \psr} \label{sec:timing} 

\begin{figure}
\begin{center}
\vspace{2mm}
\includegraphics[scale=0.46]{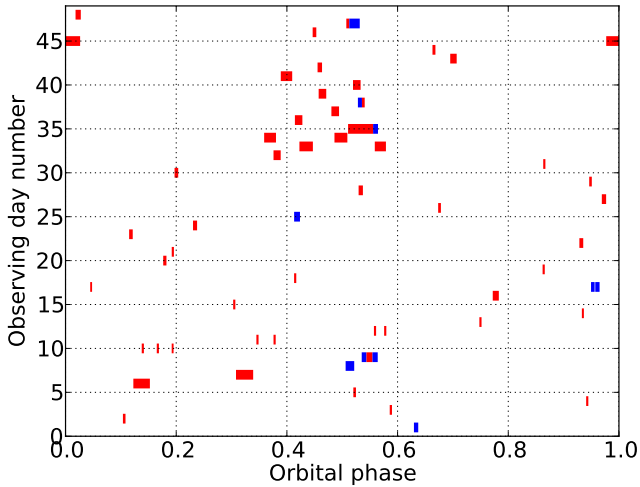}
\end{center}
\caption{\label{fig:dets}
Fifty six observations of \psr\ on 48 days during 2015 March--2016
January as a function of orbital phase $\phi_b$.  Non-detections
are represented in red, blue indicates detections.  The detection
at $\phi_b \approx 0.95$ represents two adjoining GBT observations,
at 2\,GHz and 5\,GHz respectively (see Figure~\ref{fig:profs}d--e).
The non-detections on day numbers 16 and 20 are at 2\,GHz; all other
observations are from Parkes at 1.4\,GHz.
}
\end{figure}

We had searched the location of the new pulsar once before
(Table~\ref{tab:search}), unsuccessfully. Following its discovery
we began Parkes timing observations at 1.4\,GHz.  Between 2015 March
28 and 2016 January 14 we have observed \psr\ 52 times on 46 days
for a total of 75\,hr, and have detected it for a total of 11\,hr
on seven days (Figure~\ref{fig:profs}a--c and f--i).  All of these
detections were within orbital phases $0.4 < \phi_b < 0.65$, which
we have observed for 42\,hr (see Figure~\ref{fig:dets}; throughout
this paper we use the radio phase convention, in which $\phi_b=0$
is the time of ascending node of the pulsar).  Two thirds of these
observing sessions, 85\% of the observing time, and all of the
detections, used the same PDFB4 setup employed in the discovery
observation.  The remaining timing observations were done with the
Berkeley-Parkes-Swinburne Recorder (BPSR), which in effect samples
340\,MHz of bandwidth centered on 1352\,MHz every $64\,\mu$s with
0.4\,MHz resolution \citep[for more details, see][]{kjv+10}.

For the relatively large DM of \psr\ we do not expect interstellar
scintillation to modulate the detected flux density by more than a
factor of $\sim 2$. Some, likely all, of the non-detections are
caused by eclipses, as in Figure~\ref{fig:profs} (panels c,g,h,i).
On at least two occasions (see ``kinks'' lasting for $\approx 500$\,s
near the beginning/end of the observations in Figure~\ref{fig:profs}h/i)
the pulse arrival is delayed by $\approx 0.15\,P$, presumably due
to an ephemeral clump of ionized material. This time delay corresponds
to an extra electron column $\delta\mbox{DM} \approx 0.15$\,pc\,cm$^{-3}$.

In 2015 June we observed \psr\ at the Robert C.\ Byrd Green Bank
Telescope (GBT). We did this on three days for a total of 3\,hr at
a central frequency of 2\,GHz, recording a bandwidth of 800\,MHz
using
GUPPI\footnote{\url{https://safe.nrao.edu/wiki/bin/view/CICADA/GUPPiUsersGuide}}.
One of these observations resulted in a detection
(Figure~\ref{fig:profs}d), at $\phi_b \approx 0.95$
(Figure~\ref{fig:dets}). On the same day we also did a 1\,hr
observation using the new GBT C-band receiver, centered at 4.8\,GHz
with 800\,MHz bandwidth, and detected the pulsar with an estimated
flux density of $\approx 40\,\mu$Jy (Figure~\ref{fig:profs}e).

Additionally, we observed \psr\ during 2015 April--June at Parkes
at 0.7\,GHz (once for 0.6\,hr) and 3.1\,GHz (twice for a total of
1.3\,hr), and at the GBT at 350\,MHz (three times, 0.6\,hr). Finally,
in a different Parkes survey of \fermi\ unidentified sources, we
had also observed its location at 1.4\,GHz in 2014 July (three
times, 2.1\,hr). None of these observations resulted in detections.

\begin{deluxetable}{ll}
\tablecaption{\label{tab:j1417} Parameters of \psr\ }
\tablecolumns{2}
\tablehead{
\colhead{Parameter} &
\colhead{Value}
}
\startdata
\cutinhead{Timing Parameters}
Right ascension, R.A. (J2000.0)\tablenotemark{a}
  & $14^{\rm h}17^{\rm m}30\fs604$                         \\
Declination, decl. (J2000.0)\tablenotemark{a}
  & $-44\arcdeg 02\arcmin 57\farcs37$                      \\
Spin period, $P$ (ms) & 2.6642160(4)                       \\
Dispersion measure, DM (pc\,cm$^{-3}$) &  55.00(3)         \\
Binary period, $P_b$ (d) & 5.37372(3)                      \\
Projected semi-major axis, $x_1$ (l-s) & 4.876(9)          \\
Time of ascending node, $T_{\rm asc}$ (MJD) & 57111.457(2) \\
[-5pt]
\cutinhead{Derived Parameters}
Galactic longitude, $l$ & $318\fdg86$                 \\
Galactic latitude, $b$  & $16\fdg14$                  \\
DM-derived distance, $d$ (kpc)\tablenotemark{b} & 1.6 \\
Pulsar mass function, $f_1$ (\msun) & 0.00433         
\enddata
\tablecomments{Numbers in parentheses represent uncertainties on
the last digit. See Section~\ref{sec:timing} for a description of
the fits. }
\tablenotetext{a}{Fixed at the optical position \citep{scc+15}.}
\tablenotetext{b}{From the electron density model of \citet{cl02}
(see Section~\ref{sec:dist}).}
\end{deluxetable}

There are currently too few radio detections to obtain a phase-connected
timing solution. However, the existing detections (all of which
indicated orbital acceleration) were sufficient to determine that
\psr\ is the neutron star in the \citet{scc+15} binary: it has the
same orbital period and its phase is $180\arcdeg$ shifted from that
of the optical star. In order to obtain the parameters listed in
Table~\ref{tab:j1417} we first extracted pulse times of arrival
(TOAs) from all detections. Using the
TEMPO\footnote{\url{http://tempo.sourceforge.net}} timing software
we fit each day's TOAs to a model including spin frequency and up
to two derivatives as needed. We then did a least-squares fit of a
circular orbit ($e<0.02$ according to \citealt{scc+15}) to the
resulting overall set of Doppler-shifted barycentric spin frequencies.
We also used TEMPO to directly fit a zero-eccentricity binary model
to the TOAs, allowing for an arbitrary offset between daily sets
of TOAs (i.e., the resulting solution is not phase connected). The
parameters and uncertainties given in Table~\ref{tab:j1417} encompass
the sets of values returned by these two independent fits.  Our
scant detections currently span a narrow range of binary phases
(Figure~\ref{fig:dets}), and we expect greatly improved orbital
parameters as detections accumulate; nevertheless, our values for
$P_b$ and $T_{\rm asc}$ already have 10 times the precision of those
obtained from optical observations.

\subsection{X-ray and UV Observations of \psr} \label{sec:swift} 

The \swift\ X-ray Telescope (XRT, \citealt{bhn+05}) observed \psr\
in 2015 March and June, for a total of 12.5\,ks.  The eight
observations in March spanned slightly more than one binary orbit
(Table~\ref{tab:swift}).  We processed all \swift\ observations
using FTOOLS version
6.16\footnote{\url{http://heasarc.gsfc.nasa.gov/ftools}}.  The
0.3--10\,keV spectrum and light curve from the XRT photon counting
mode data were obtained with XSELECT. The source events were extracted
from a circular aperture of radius 20 pixels ($47''$), while the
background was taken from an annulus with an inner radius of 60
pixels and an outer radius of 110 pixels.

\begin{deluxetable*}{llccccc}
\tabletypesize{\small}
\tablewidth{0pt}
\tablecaption{\label{tab:swift} \swift\ XRT and UVOT Observations of \psr\ }
\tablecolumns{7}
\tablehead{
\colhead{ObsID}                         &       
\colhead{Date}                          & 
\colhead{Exposure Time}                 & 
\colhead{Binary Phase\tablenotemark{a}} &
\colhead{XRT Count Rate}                &   
\colhead{UVOT Filter}                   &
\colhead{Magnitude\tablenotemark{b}}    \\
\colhead{}                                     &
\colhead{(UT)}                                 &
\colhead{(ks)}                                 &
\colhead{($\phi_b$)}                           &
\colhead{(0.3--10\,keV) ($10^{-2}$\,s$^{-1}$)} &
\colhead{}                                     &
\colhead{}
}
\startdata
00084749001 & 2015 Mar 13 & 0.4 & 0.89 & $1.8^{+0.9}_{-0.7}$ & $uvm2$ & $>18.60$ \\
00084749002 & 2015 Mar 14 & 1.8 & 0.94 & $1.2^{+0.4}_{-0.3}$ & $uvw1$ & $ 18.98 \pm 0.11 \pm 0.03$ \\
00084749004 & 2015 Mar 15 & 1.6 & 0.33 & $1.1\pm0.3$         & $u$    & $ 18.22 \pm 0.05 \pm 0.02$ \\
00084749005 & 2015 Mar 16 & 1.8 & 0.41 & $2.35\pm0.45$       & $uvw2$ & $>20.26$ \\
00084749006 & 2015 Mar 17 & 0.5 & 0.55 & $4.0\pm1.0$         & $uvm2$ & $>19.93$ \\
00084749007 & 2015 Mar 18 & 1.4 & 0.78 & $0.8^{+0.4}_{-0.3}$ & $uvw1$ & $>18.22$ \\
00084749008 & 2015 Mar 19 & 1.4 & 0.95 & $1.3\pm0.4$         & $u$    & $ 17.85 \pm 0.04 \pm 0.02$ \\
00084749009 & 2015 Mar 20 & 1.7 & 0.09 & $1.8\pm0.4$         & $uvw2$ & $>19.84$ \\
00084749010 & 2015 Jun 18 & 2.1 & 0.82 & $1.7\pm0.3$         & $uvw1$ & $ 19.28 \pm 0.13 \pm 0.03$
\enddata
\tablecomments{All uncertainties represent 90\% confidence levels.}
\tablenotetext{a}{Phase 0.75 corresponds to companion superior
conjunction. These $\phi_b$ values are based on the radio ephemeris
of Table~\ref{tab:j1417}. }
\tablenotetext{b}{For measured magnitudes (in the Vega system,
\citealt{blh+11}), the first uncertainty is statistical and the
second is systematic. }
\end{deluxetable*}

The X-ray count rates appear to show variability, with a maximum
at binary phase $\phi_b = 0.55$. However, the probability that the
observed counts arise from a constant flux distribution is 2.4\%
(determined from a $\chi^2$ fit of a model with a constant count
rate to the X-ray light curve), so we cannot rule out the null
hypothesis that the flux is constant.  The combined time-averaged
X-ray spectrum is well-fitted using XSPEC by an absorbed power-law
with photon index $\Gamma=1.59^{+0.35}_{-0.20}$, column density
$N_{\rm H} = 2.2^{+1.7}_{-1.3}\times 10^{21}$\,cm$^{-2}$, and an
unabsorbed 0.3--10\,keV flux of $(8.7\pm1.5) \times
10^{-13}$\,erg\,cm$^{-2}$\,s$^{-1}$ (all uncertainties in this
section represent 90\% confidence levels).  This is consistent with
the spectrum measured from a 2011 \chandra\ observation \citep{scc+15},
and corresponds to an X-ray luminosity of $\approx 3\times
10^{32}\,(d/1.6\,\mbox{kpc})^2$\,erg\,s$^{-1}$, which we scale
according to the DM-derived distance (see Section~\ref{sec:dist}
for a discussion of this).

Simultaneously with the XRT, \psr\ was observed with the \swift\
ultraviolet and optical telescope (UVOT, \citealt{rkm+05}), through
a variety of filters. The resulting magnitudes and upper limits are
also presented in Table~\ref{tab:swift}.  The UVOT photometric
measurements were obtained using the {\tt uvotsource} command in
FTOOLS. The source flux was obtained from a $5''$ circle, while the
background was taken from a source-free circular region of radius
$20''$.  There are two detections in $u$ and two in $uvw1$.  The
remaining UV observations yield only upper limits.  The implications
of these will be discussed in Section~\ref{sec:disk}.

\section{Discussion} \label{sec:disc} 

\subsection{What is the Distance to \psr?} \label{sec:dist}

\citet{scc+15} estimate that the \psr\ system lies at a distance
of $d= 4.4$\,kpc (with a likely uncertainty of at least 20\%) based
on the magnitude and optical spectrum of the companion, and assuming
that it fills its Roche lobe.  Given the poor pulsar detectability
(generally low signal-to-noise ratio and frequent eclipses), it is
unlikely that a parallax will be measured either interferometrically
or through timing observations.  But we can estimate the distance
from the radio pulsar DM. The \citet{cl02} electron distribution
model (NE2001) gives $d_{\rm NE2001} = 1.6$\,kpc, while the previously
most used electron density model gives $d_{\rm TC93} = 3.0$\,kpc
\citep{tc93}.

We favor the NE2001 distance because TC93 generally over-predicts
distances and fails to account for the DMs of about 10\% of pulsars
\citep[][and unpublished reanalysis of the ATNF pulsar catalog]{cl02}.
Of 66 pulsars with parallax
measurements\footnote{\url{http://www.astro.cornell.edu/research/parallax}},
17 have DMs between 35 and 75\,pc\,cm$^{-3}$ (compared to
55\,pc\,cm$^{-3}$ for \psr). Of these, 10 have NE2001-estimated
distances that are consistent to within 20\% of the parallax distance
ranges (only one of the 10 had a well constrained distance used to
construct NE2001).  The remaining seven objects have distances that
disagree by more than 20\%, in some cases by a factor of 2. Four
have NE2001 distances larger than the parallax distances and the
opposite is the case for the other three pulsars.

We conclude that there is no identifiable distance bias in NE2001
(as there is for TC93) and that the NE2001 distance is the best
provisional DM-based distance estimate.  However, due to possible
variations in the scale height of the electron density distribution,
there is considerable uncertainty in determining DM distances to
pulsars at high Galactic latitude when $\mbox{DM}\,\sin |b|$ is
comparable to its maximum value (see Figure~1 of \citealt{cl03}).
\psr\ falls in this uncertain regime.  Also, a recent study reports
new MSP parallax distances that in some cases differ from the NE2001
distances by a factor of 2 \citep{mnf+16}.  Therefore, we do not
exclude a larger distance such as that derived by \citet{scc+15}.
The conflict between these independent distance measurements,
4.4\,kpc and 1.6\,kpc, extends to the assumptions made by \citet{scc+15}
in modeling the inclination angle of the system and the masses,
because the star would only fill a fraction of the Roche lobe if
at the smaller DM-based distance.  These issues are discussed in
Section~\ref{sec:masses}.

\subsection{Is There an Accretion Disk?} \label{sec:disk}

\citet{scc+15} concluded that an accretion disk is present, primarily
because of the double-peaked H$\alpha$ emission line, but also
because the X-ray luminosity at their preferred distance of 4.4\,kpc
is consistent with those of transitional MSPs in their disk states.
The detection of radio pulsations from \psr\ brings that interpretation
into question because the transitional MSPs PSR~J1023+0038
\citep{bab+15} and XSS~J12270$-$4859 \citep{hsc+11} show no evidence
for radio pulsations in their accreting states.  Along with the
detection of accretion-powered X-ray pulsations in both systems
\citep{abp+15,pdb+15}, this is an indication that the radio pulsar
mechanism of these two MSPs is completely quenched by the accretion
flow \citep[see also][and references therein]{cs08}.

If the lower DM distance is adopted, the X-ray luminosity of $\approx
3\times 10^{32}\,(d/1.6\,\mbox{kpc})^2$\,erg\,s$^{-1}$
(Section~\ref{sec:swift}) is comparable to $\sim10^{32}$\,erg\,s$^{-1}$
measured for the nearby redback PSR~J1723$-$2837 \citep{bec+14} and
lower than the average accreting luminosity of the transitional
MSPs \citep[e.g.,][]{lin14}.  Non-thermal X-rays typically observed
in redbacks seems to originate from an intra-binary shock driven
by the interaction of the pulsar wind and matter from the companion
\citep{at93}.  However, the uncertain distance to \psr\
(Section~\ref{sec:dist}) and possible X-ray variability \citep[][and
Section~\ref{sec:swift}]{scc+15} make existing X-ray observations
an unreliable discriminator for the presence or absence of accretion.

Even if no accretion flow is reaching the neutron star and the
X-rays are not accretion powered, a disk that is truncated at the
magnetospheric radius or beyond could still radiate in the ultraviolet
and visible.  However, the optical and UV continuum data disfavor
an accretion disk.  While \citet{scc+15} measured a $V_0$ magnitude
(corrected for extinction) spanning 15.64--15.91, \psr\ is much
fainter in $u$ and especially in the UV filters (Table~\ref{tab:swift}).
Correcting the \swift\ $u$ magnitudes for an extinction $A_u=0.50$
\citep{sf11}, we find $u_0=17.35$ and 17.72 at $\phi_b=0.95$ and
0.33, respectively.  Then $1.7\le u_0-V_0\le 1.9$, which is in the
correct range for giant stars of spectral type G8--K0, which have
$1.5\le U_0-V_0\le1.9$ \citep{pic98}.  This is the type inferred
by \citet{scc+15} from the optical spectrum of the companion.  Thus,
there is no evidence for extra light in the ultraviolet coming from
an accretion disk.  Also, the ultraviolet flux detections, being
higher at $\phi_b=0.95$ than at $\phi_b=0.33$, and faintest at
$\phi_b=0.82$, are consistent in magnitude and phase with the
observed modulation due to ellipsoidal variations that dominate the
optical light curves in \citet{scc+15}.

However, since the companion star is likely a giant or subgiant
(see below), it is brighter than the redback companions and could
mask an accretion-disk component.  We can estimate an upper limit
on a mass-transfer rate under the conservative assumption that
$<50\%$ of the minimum observed $u$-band flux is contributed by a
standard blackbody disk \citep{fkr02} that is truncated at the
pulsar corotation radius, $r_{\rm co}=36$\,km for a 2\,\msun\ neutron
star. For an inclination angle of $58\arcdeg$ (see
Section~\ref{sec:masses}) and a distance of 1.6\,kpc, the limit is
$\dot m<4\times10^{14}$\,g\,s$^{-1}$.  The $u$-band luminosity of
a disk is barely changed even if it is truncated at the light
cylinder radius, $r_{\rm lc}=127$\,km for \psr.

This can be compared with $\dot m \approx1\times10^{15}$\,g\,s$^{-1}$
estimated by \citet{pt15} for the demonstrably accreting transitional
objects PSR~J1023+0038 and XSS J12270$-$4859 using similar assumptions.
And if the distance to \psr\ is actually 4.4\,kpc, then it could
have a similar accretion rate as the transitional pulsars.  Note
that, in all of these cases, a disk with such a low accretion rate
cannot actually reach the corotation radius, and most of the matter
is propelled away \citep{pt15}.  This leaves the possibility that
the H$\alpha$ could be emitted either from the outer disk or in a
wind.

The current properties of \psr\ appear to be most reminiscent of
the eclipsing radio MSP J1740$-$5340 in the globular cluster NGC~6397
\citep{dpm+01}, which is bound to a ``red straggler''/sub-subgiant
companion in a 1.3 day orbit. However this system was likely formed
in an exchange interaction in the dense cluster environment
\citep{ov03}, with an evolutionary path substantially different
from that of \psr.  The bolometric luminosity of the \psr\ companion,
determined by \cite{scc+15} for a distance of 4.4\,kpc, is reduced
by a factor of 7.6 at the DM-derived 1.6\,kpc, making it comparable
to that of subgiants.  PSR~J1740$-$5340 exhibits radio eclipses
around superior conjunction for 40\% of the orbit, as well as random
DM variations at all phases.  Its X-ray spectrum is non-thermal
(with $\Gamma=1.7$) and possibly modulated at the binary period
\citep{bvh+10}, suggestive of an intra-binary shock.  Perhaps most
importantly, PSR~J1740$-$5340 also exhibits a prominent H$\alpha$
emission line with complex morphology and strong orbital phase
dependence \citep{sgf+03}, which can be interpreted as being produced
in part by a wind from the secondary star that is driven out of the
binary by the pulsar wind.  The clear evidence for an active
rotation-powered radio pulsar suggests that \psr\ is presently in
the so-called ``radio ejection'' regime \citep{rst89a,bdb02} in
which any Roche lobe overflow through the L1 point cannot overcome
the barrier imposed by the pulsar wind pressure.  A larger orbit
actually favors this scenario according to \citet{bdb02}.

One might also consider the possibility that the system transitions
rapidly between accreting and non-accreting states.  But H$\alpha$
emission, as evidence for accretion, was detected by \citet{scc+15}
in more than one dozen observations spanning 1\,yr through 2015
February, while the first detection of radio pulsations, ruling out
accretion onto the neutron star, occurred only one month later
(Section~\ref{sec:search}).  Given all the available evidence, we
favor an interpretation in which the \psr\ system was not accreting
at the time of the optical and radio observations.

\subsection{Component Masses} \label{sec:masses}

If the mass ratio and the inclination angle of the system can be
determined with precision, then the mass of the neutron star can
also be measured. This is of particular interest because \citet{scc+15}
concluded that the neutron star is massive, $M_1=1.97\pm0.15$\,\msun.
Assuming a tidally locked, Roche-lobe filling secondary, they
obtained a mass ratio $q \equiv M_2/M_1 = 0.18 \pm 0.01$ from a
measurement of the projected rotational velocity of the secondary
(which has unspecified systematic uncertainties) and its radial
velocity semi-amplitude $K_2 = 115.7 \pm 1.1$\,km\,s$^{-1}$.  They
also assumed a Roche-lobe filling star to fit the optical light
curves for the orbital inclination angle using an ellipsoidal model,
finding $i = 58\arcdeg\pm2\arcdeg$.

\begin{figure}
\begin{center}
\includegraphics[scale=0.70,angle=270]{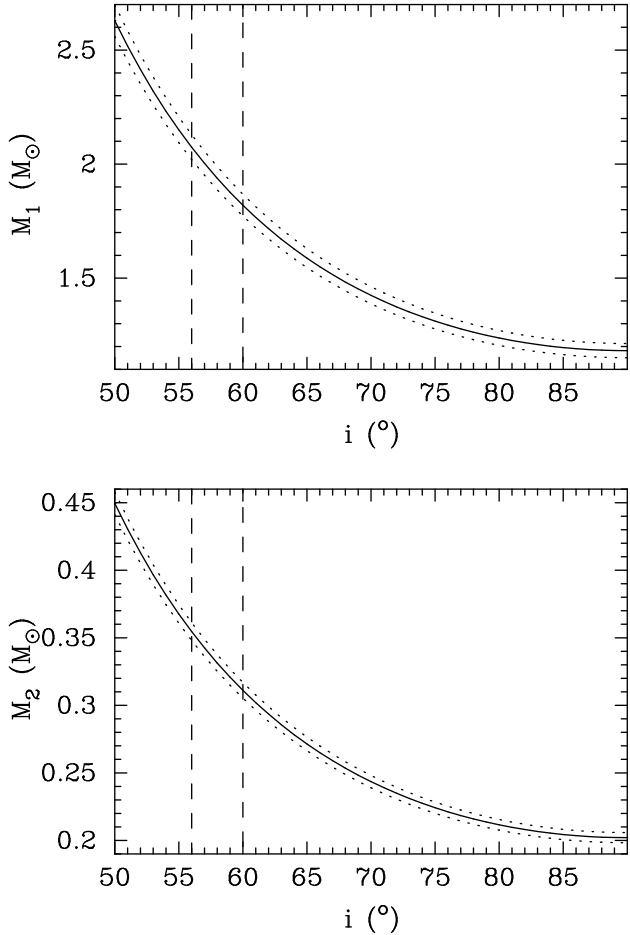}
\end{center}
\caption{\label{fig:mass}
Masses of the neutron star (top) and companion (bottom) as a function
of inclination angle $i$.  The dotted curves allow for the uncertainties
in the measured orbital velocities of the two stars, as quoted in
the text.  \citet{scc+15} modeled $i$ as $58\arcdeg\pm2\arcdeg$
(vertical dashed lines).  But $i$ should be larger if the companion
is not filling its Roche lobe.
}
\end{figure}

Using the pulsar projected semi-major axis $x_1$ (Table~\ref{tab:j1417})
we infer that the pulsar radial velocity semi-amplitude is
19.8\,km\,s$^{-1}$. Along with $K_2$ this implies $q = 0.171 \pm
0.002$, limited by the precision on $K_2$.  Figure~\ref{fig:mass}
shows the masses of the neutron star and companion as a function
of inclination angle, allowing for the uncertainties on $K_2$ and
$x_1$.  For $56\arcdeg\leq i \leq60\arcdeg$, $M_1$ ranges over
1.77--2.13\,\msun, and $M_2 = 0.33\pm0.03$\,\msun.  However, if
located at the smaller DM-based distance, the radius of the companion
is only 0.36 times the Roche-lobe radius; its tidal distortion is
reduced, and a larger $i$ would be necessary to fit the ellipsoidal
modulation, resulting in smaller masses.  The minimum allowed masses
(for $i = 90\arcdeg$) are $M_2\geq0.20$\,\msun\ and $M_1\geq1.15$\,\msun.
The Roche lobe radius decreases only slightly, from $4.1$\,\rsun\
at $i = 58\arcdeg$ to $3.5$\,\rsun\ at $i = 90\arcdeg$ according
to the formula of \citet{egg83}, so a higher inclination at the
smaller distance would not allow the companion to contact the Roche
lobe.

\subsection{No Photospheric Heating} \label{sec:heating}

The optical light curves \citep{scc+15}, dominated as they are by
ellipsoidal modulation, show no evidence for heating of the companion's
photosphere by the pulsar wind.  This is probably because the
intrinsic luminosity of the giant/subgiant dominates.  We can place
an upper limit on the spin-down luminosity $\dot E$ of the pulsar,
or more precisely, on the product of $\dot E$ and an efficiency
factor $\eta$ that includes the physics of reprocessing and any
anisotropy of the pulsar wind.  A heating effect can be approximated
using $\eta\dot E R_2^2/4a^2=2\pi R_2^2\sigma(T_h^4-T_{\rm eff}^4)$,
where $T_{\rm eff}$ is the intrinsic temperature of the unheated
star, $T_h$ is the average temperature of the side facing the pulsar,
$R_2$ is the radius of the companion, and $a$ is the orbital
separation.

We use $T_{\rm eff}=5000$\,K from the optical spectrum of \citet{scc+15},
and assume an upper limit of 5200\,K for $T_h$ based on the absence
of a heating signature at the level of 0.1--0.2 magnitudes.  For
$i=58\arcdeg$, the orbital separation is $a=17$\,\rsun, and the
limit on $\eta \dot E$ is $2.1\times10^{35}$\,erg\,s$^{-1}$.  If
instead we assume as a maximum $i=72\arcdeg$ (for which $M_1=1.4$\,\msun),
then $a=15.2$\,\rsun\ and $\eta \dot E<1.7\times10^{35}$\,erg\,s$^{-1}$.
Neither of these limits are very restrictive, as $\eta$ is likely
to be less than unity.  These limits can also accommodate the
gamma-ray luminosity of \fgl, $3\times10^{34}$\,erg\,s$^{-1}$ at a
distance of 4.4\,kpc (see Section~\ref{sec:conc}).

\subsection{Evolutionary Track} \label{sec:evol}

With a binary period of 5.4 days, \psr\ is on the evolutionary track
of low-to-intermediate mass companions that started mass transfer
onto a neutron star after they left the main sequence \citep{prp02b},
and have increasing periods.  This distinguishes them from the black
widows, redbacks, and other low-mass X-ray binaries that have binary
periods $\leq1$\,day, began mass transfer on the main sequence, and
evolve toward shorter periods.  For 1\,\msun\ secondaries, these
classes are separated by a bifurcation period of $\approx18$\,hr
at the onset of mass transfer.

With $M_2 \approx 0.3$\,\msun, the companion of \psr\ must have
already lost most of its initial mass, only a fraction of which,
accreting onto the neutron star, is sufficient to spin it up to
2.66\,ms.  In the models of \citet{prp02b}, \psr\ is not far from
the ends of the tracks that terminate when the binary period is in
the range 6--20 days, and the companion is a He white dwarf of
0.2--0.3\,\msun.  The neutron stars in these calculations, assumed
to begin with $1.4$\,\msun, end in the range 1.8--2.2\,\msun.
Exactly how far \psr\ is along this evolution is difficult to
determine, since its companion may or may not be filling its Roche
lobe, may or may not be finished accreting onto the neutron star,
and may be losing mass mostly through ablation by the pulsar wind,
a process that is difficult to model.

\section{Conclusions} \label{sec:conc}

We have discovered a 2.66\,ms pulsar in orbit around the 5.4 day
optical variable discovered by \citet{scc+15} in the error ellipse
of the high-energy gamma-ray source \fgl.  The orbital dynamics
require that the companion mass is $>0.2$\,\msun.  The presence of
radio pulsations suggests that the neutron star is not currently
accreting (see Section~\ref{sec:disk}).  The pulsar is only detected
a fraction of the time it is observed (although few observations
exist at orbital phases $0.65 < \phi_b < 0.9$; Figure~\ref{fig:dets}),
and is likely eclipsed for more than half the orbit by outflowing
material from the companion; this has so far prevented us from
obtaining a phase-connected timing solution.

The DM distance of 1.6\,kpc is in conflict with the original estimate
of $d=4.4$\,kpc, which was based on the assumption that the companion
fills its Roche lobe.  This ambiguity then affects most of the
remaining interpretation, including the precise evolutionary state
of the system, whether it currently has an accretion disk, and the
masses, which depend on the modeled inclination angle.  The current
estimate of $1.77\leq M_1\leq2.13$\,\msun\ should really be regarded
as an upper limit.  There are several claims of massive neutron
stars in redbacks and black widows. Modeling these systems, however,
is difficult, especially in the presence of significant heating of
the secondary by the pulsar \citep[see, e.g.,][]{rgfk15}.  So far
there is no evidence for heating in the \psr\ system, which makes
it a good candidate for a precise mass measurement.  Further optical
observations and modeling of ellipsoidal variations may better
determine the inclination angle, and thus the value of $M_1$, whose
uncertainty is currently dominated by the systematic uncertainty
on $i$.

The gamma-ray luminosity of \fgl\ is $L_{\gamma}=3.7\times
10^{33}\,(d/1.6\,\mbox{kpc})^2$\,erg\,s$^{-1}$ \citep[see][]{aaa+15},
for isotropic emission.  At the DM-derived distance this could be
accounted for by magnetospheric emission from an MSP with a typical
spin-down luminosity $\dot E\sim10\,L_{\gamma}$.  At the larger
distance favored by \citet{scc+15}, \psr\ would be among the three
most luminous MSPs \citep[see][]{aaa+13}, and might require substantial
gamma-ray emission from an intra-binary shock, or a higher than
average spin-down luminosity.

A phase-connected timing solution for \psr\ will yield its $\dot
E$ and will allow a search for gamma-ray pulsations.  The spin-down
luminosity is a fundamental quantity that limits the X-ray and
gamma-ray luminosities, and the possible rate of ablation of the
companion and of the inner accretion disk if any exists.  In addition,
a phase-connected solution will place useful constraints on the
orbital eccentricity. These quantities should lead to an improved
understanding of this unusual system. Finally, the great difficulty
we have in detecting the radio pulsar is a reminder that there could
be a substantial population of such long-period systems awaiting
discovery, which would represent the late phases, until now unobserved,
in the formation of typical MSP binaries.

\acknowledgements

We are grateful to the ATNF Electronics group led by Warwick Wilson
for their work with the digital filterbanks. We thank Ryan Shannon,
Dick Manchester, and George Hobbs for observing assistance, and
Lawrence Toomey for help with accessing archival data. Phil Edwards
accommodated our stringent observing constraints with grace and
efficiency, for which we are most thankful. We acknowledge stimulating
feedback from Thomas Tauris and Roger Romani. The Parkes Observatory
is part of the Australia Telescope, which is funded by the Commonwealth
of Australia for operation as a National Facility managed by CSIRO.
The National Radio Astronomy Observatory is a facility of the
National Science Foundation operated under cooperative agreement
by Associated Universities, Inc. We acknowledge the use of public
data from the \swift\ data archive. Work by P.\ Ray is supported
by the NASA \fermi\ Guest Investigator program.

{\em Facilities:}  \facility{Parkes (PDFB, BPSR)}, \facility{GBT
(GUPPI)}, \facility{Swift (XRT, UVOT)}

\end{document}